\documentclass[referee]{aastex} % for a referee version

\newcommand\lsim{\lower0.5ex\hbox{$\; \buildrel < \over \sim \;$}}
\shortauthors{Li et al.}

\usepackage{longtable}
\usepackage{color}
\usepackage{float}

\begin{document}

\title{Observing Kelvin$-$Helmholtz instability in solar blowout jet}

\author{Xiaohong Li\altaffilmark{1,2}, Jun Zhang\altaffilmark{1,2}, Shuhong Yang\altaffilmark{1,2},
Yijun Hou\altaffilmark{1,2} \& Robert Erd\'elyi\altaffilmark{3,4}}

\altaffiltext{1}{CAS Key Laboratory of Solar Activity, National
Astronomical Observatories, Chinese Academy of Sciences, Beijing
100101, China}

\altaffiltext{2}{School of Astronomy and Space Science, University
of Chinese Academy of Sciences, Beijing 100049, China}

\altaffiltext{3}{Solar Physics and Space Plasma Research Centre, School
of Mathematics and Statistics, University of Sheffield, Hicks Building,
Hounsfield Road, Sheffield S3 7RH, UK}

\altaffiltext{4}{Department of Astronomy, E\"otv\"os Lorand University,
P\'azm\'any P\'eter s\'et\'any 1/A, Budapest, H-1117, Hungary}

{\bf Kelvin$-$Helmholtz instability (KHI) is a basic physical process in fluids and magnetized plasmas,
with applications successfully modelling e.g. exponentially growing instabilities observed at magnetospheric
and heliospheric boundaries, in the solar or Earth's atmosphere and within astrophysical jets. Here,
we report the discovery of the KHI in solar blowout jets and analyse the detailed evolution by employing
high-resolution data from the Interface Region Imaging Spectrograph (IRIS) satellite launched
in 2013. The particular jet we focus on is rooted in the surrounding penumbra of the main negative polarity sunspot of
Active Region 12365, where the main body of the jet is a super-penumbral structure. At its maximum, the jet has a
length of 90 Mm, a width of 19.7 Mm, and its density is about 40 times higher than its surroundings.
During the evolution of the jet, a cavity appears near the base of the jet, and bi-directional flows
originated from the top and bottom of the cavity start to develop, indicating that magnetic reconnection
takes place around the cavity. Two upward flows pass along the left boundary of the jet successively.
Next, KHI develops due to a strong velocity shear ($\sim$ 204 km s$^{-1}$) between these two flows, and
subsequently the smooth left boundary exhibits a sawtooth pattern, evidencing the onset of the instability.}

KHI, as proposed by Lord Kelvin in 1871$^{1}$ and Helmholtz in 1868$^{2}$, occurs when two parallel
fluids with different velocities flow alongside each other, with a shear exceeding a
critical value$^{3}$. KHI has been shown to be important to understand a considerable number of
astrophysical and space physical phenomena such as the dynamic structure at cometary tails$^{4}$,
relativistic outflows and oscillations in astrophysical jets$^{5,6}$, the merger of neutron
star systems$^{7}$ or the transfer of energy and momentum between the solar wind and solar
system's planetary magnetospheres$^{8-11}$. KHI may also play a very important role in
plasma heating as it develops small scales therefore enhancing the dissipative processes,
such as turbulent viscosity$^{12}$. High-resolution observations with modern space
instruments enable us to detect the KHI in the Sun, e.g. at the boundaries of lager-scale coronal
mass ejections (CMEs)$^{13-16}$ or within fine-scale structures (e.g. filaments$^{17}$)
in the solar corona$^{18}$.

Solar jets$^{19}$ occur over a wide range of scales in the solar atmosphere from spicules $^{20}$,
anemone jets$^{21}$, to network jets$^{22}$, X-ray jets$^{23,24}$, etc. Often they are localised plasma
eruptions likely to develop at e.g. magnetic reconnection areas where new emerging flux reconnects with
pre-existing fields which have opposite magnetic polarities$^{25}$. A blowout model$^{26}$ has been
proposed to explain the formation of jets, i.e. the strong shear and twist of the magnetic field in
the core of the jet's arch drive an ejective eruption, producing blowout jets$^{27,28}$.
We have observed KHI in several solar jets, and here we establish the most distinct and most detailed
example observed by IRIS$^{29}$.

The IRIS spacecraft yields simultaneous spectral and imaging observations of the solar atmosphere.
On 12 June 2015, there were 8 active regions (ARs) on the solar disk. One of them, AR 12365, was
located southeast of a trans-equatorial coronal hole. At the boundary of the main sunspot of this AR,
several jets occurred successively. IRIS observed AR 12365 with a pixel size of 0.$\arcsec$33 and a
cadence of 3 seconds. The high temporal resolution enables us to study the evolution of the jets in
unprecedented details. We also employ the associated magnetograms taken by the Helioseismic and Magnetic
Imager (HMI)$^{30}$ and the concurrent multi-wavelength images sampled by the Atmospheric Imaging
Assembly (AIA)$^{31}$ on-board the Solar Dynamics Observatory (SDO)$^{32}$ to examine the magnetic
and velocity field evolution of the AR, in particular, the dynamics of the jets present therein.

Figure 1 shows the photospheric magnetic field of the AR and the jet in EUV observed by SDO on 12 June 2015.
The positive magnetic flux (visualised by white color) emerged at the western boundary of the main negative
polarity (dark colored) sunspot of the AR (Supplementary Video 1). At the magnetic reconnection region, below
which the emerging positive flux cancelled out with the pre-existing negative fields, a jet appeared at 19:24 UT,
and reached its emission peak at about 19:32 UT. A filament, which was anchored in the penumbra of the main
sunspot with negative polarity, was involved in the jet and filament material was ejected upwards. This process
was clearly detected in both higher and lower temperature wavelengths, e.g., 131 {\AA}, 171 {\AA} and 304 {\AA}
(Fig. 1c$-$e and Supplementary Video 2). The footpoint of the jet is shown in
Fig. 1d (red circle) and is also overlaid on the corresponding magnetogram (Fig. 1b). There were concurrent
brightenings in the jet tail as shown in Fig. 2a (see the green boxes at 19:29:56 UT and 19:33:29 UT).
During its evolution, the jet expanded with the fastest broadening taking place from 19:30 UT to 19:35 UT,
reaching its widest width of 19.7 Mm (Fig. 2b). A cavity structure above the jet's base appeared
during this broading period (denoted by blue dotted circle in Fig. 2a and Supplementary Video 3).
Meanwhile, there were rapid bi-directional flows detected from the cavity. The velocities of the upflows were
about 224 $-$ 476 km s$^{-1}$ and the downflows velocities were 88 $-$ 153 km s$^{-1}$ (Fig. 2c).
Such bi-directional motion was also found in a rather different context by e.g. 33.
Comparing the temperature map (Fig. 1f) derived from AIA observations with IRIS images (Fig. 2a),
we find that the temperature near the cavity was much higher, nearly 15.4 MK. All these rather detailed
features suggest that magnetic reconnection took place around the cavity, and that magnetic energy
was released resulting in changes of the magnetic topological structure near the jet's base.
After the peak stage of the jet formation, there were backflows from the tail of the jet, with
velocities of $\sim$ 90 km s$^{-1}$ (see Fig. 2d, denoted by v4$-$v6) in the direction perpendicular
to the line of sight. The Doppler blueshift velocity of the backflows measured by the spectra data is
about 10.3 km s$^{-1}$ (see Supplementary Fig. 1). The jet also displayed clockwise rotation seen
from the jet's base (see Methods and Supplementary Fig. 2 for details).

Let us now focus on the formation of the KHI. Figure 3 and Supplementary Video 4 show the development
of the KHI in the jet. The left boundary of the jet was very smooth (Fig. 3a) initially.
At about 19:33:15 UT (see Supplementary Video 4), a thin layer of bulk motion ``F1"
with an estimated thickness of 630 km, originating from the jet's base, passed along the left side
of the jet for about an extent of 20 Mm. We employ the method of tracking the movement of a small bright
point in ``F1" and determine the velocity of ``F1" (see Methods for details) to be about 204 km s$^{-1}$ (Fig. 3b).
Some 80 seconds later, another thin flow strip, labelled as ``F2" with a width of 460 km, also
passed along the left side of the jet, on the right of and close to ``F1." The ``F2" velocity was
estimated 264 km s$^{-1}$ (Fig. 3c), meanwhile ``F1" slowed down to 60 km s$^{-1}$. Thus, a strong
velocity gradient (i.e. shear) of about 204 km s$^{-1}$ has formed between these two flow streams.
Assuming that the jet consists of an ensemble of thin magnetic flux tubes$^{34}$, we draw a cartoon to
describe the development of the KHI (see Fig. 4). As shown in Fig. 4a and b, the plasmas in
the flux tubes, with different densities, flow alongside each other with different velocities,
causing the onset of the KHI. Vortices form at the interface between these two flux
tubes, and the boundary of the flux tubes become distorted (see Fig. 4c-f).
Based on the theoretical consideration (see Methods for details), we estimate that the KHI
could happen with a velocity difference of 204 km s$^{-1}$. In deed, the KHI
did manifest as the left boundary of the jet became distorted, which is demonstrated
in Fig. 3d$-$f. The boundary turned into a sawtooth-shape as shown in Fig. 3f, with a maximum distortion
of 1.6 Mm. This vortex-shaped boundary surface is an important feature when the KHI takes place$^{35}$.
The growth rate of the KHI is measured to be approximately 0.063, the same order of magnitude as
the theoretical value (see the explanations in Methods and Supplementary Fig. 3 for details).
Considering the distance between two vortices as the characteristic length scale for the
associated wavelength ($\lambda$ $\sim$ 5000 km), and half of
the flow widths as the estimated boundary layer thickness (a $\sim$ 545 km), then the wavelength satisfies
the condition of the fastest growing KH mode given by $\lambda$ = (2$-$4) $\times$ $\pi$ $\times$ a$^{36}$,
that is another verification for KHI really taking place.
During the evolution of the KHI, the temperature of the plasma where the KHI occurred increased by approximately
2 MK (see Supplementary Fig. 4 and Methods). When the KHI is just manifested, we find that the material at the
left boundary of the jet became rotating in the direction contrary to that of the jet (Supplementary Fig. 5).
By tracking some small bright points, we have determined the rotation diameter $\sim$ 1800 km, with
an angular velocity of $18^{\circ}$ s$^{-1}$ and projected velocity of 180 km s$^{-1}$.

Here, we clearly see and report that two upward flows, with a strong velocity shear of 204 km s$^{-1}$,
travelling parallel to each other, drive the onset of the KHI in a blow-out jet in the solar atmosphere.
The high spatio-temporal resolution capability from IRIS
enable us to actually detect the developing process (80 seconds, see Fig. 3b and 3e) and the distortion
scale ($\leqslant$ 1.6 Mm, $\sim$ 2 arcsec) of the KHI. The formed jet occurs in a region
possessing stronger magnetic fields, while the solar wind and solar filaments in which the KHI has previously been
detected have considerably weaker magnetic fields. Our finding extends evidently the range where the KHI
takes place in the solar atmosphere to much smaller scales, and implies that the KHI may be rather ubiquitous
on the Sun in the presence of jets. This latter point is a very important one, as the smaller the scales
where the KHI may take place, the more important this instability may be for the heating of the localised
plasma$^{37}$. So far, the lack of evidence of small-scale KHI vortices may be related to the
insufficient resolution. Considering that the KHI is undisputedly an important aspect in (magneto)hydrodynamics,
so the study of KHI is conducive not only for understanding solar activities, but also to comprehending
fundamental physical processes in fluids and magnetic plasmas.
Theoretically, there may be also other possibilities, for example, the sawtooth boundary is the
manifestation of blob formation or LOS effects of twist structures, but these explanations would require
more details to support them.

\clearpage

\begin{figure}
\centering
\includegraphics
[bb=100 160 490 680,clip,angle=0,scale=0.85]{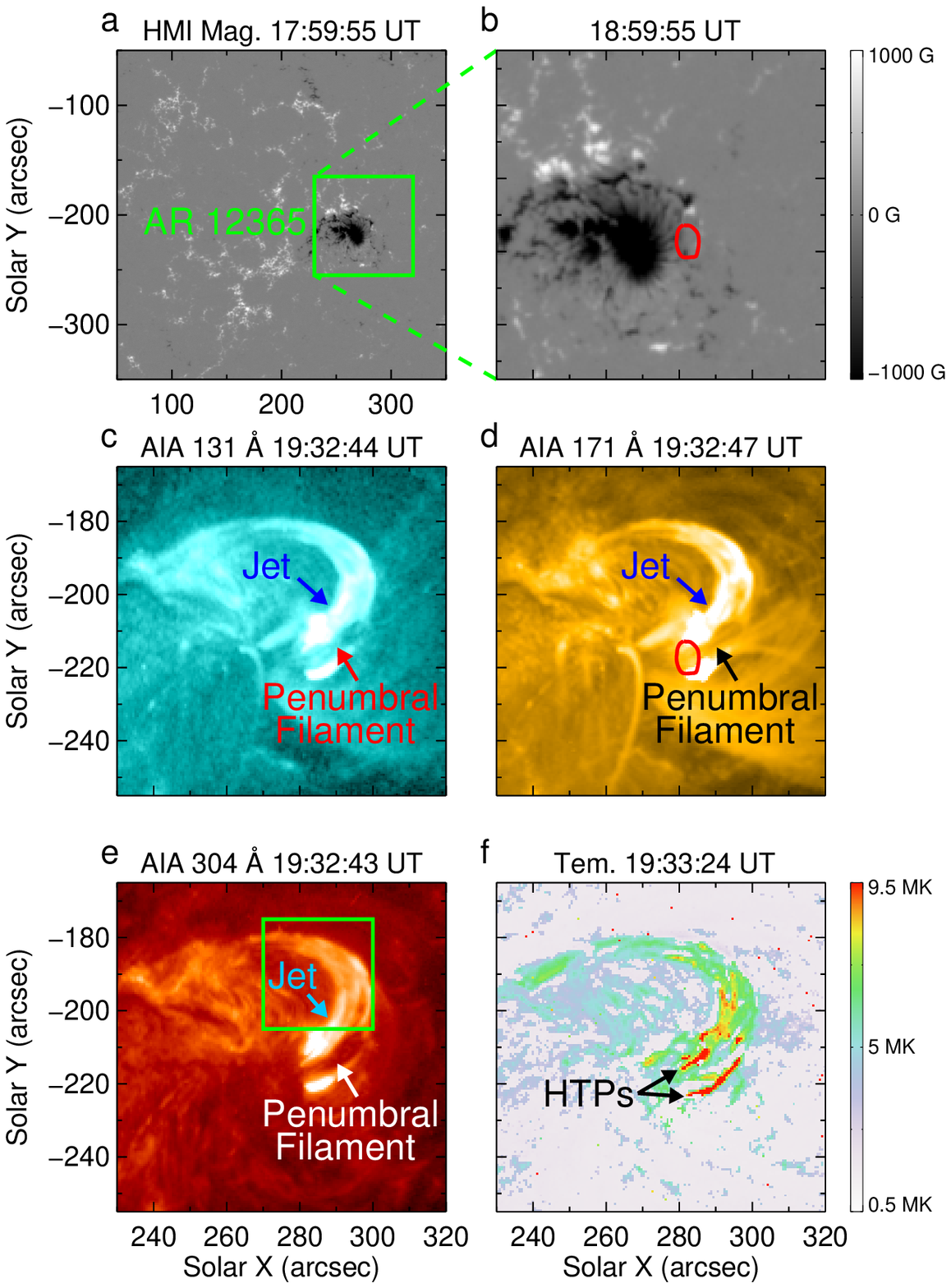}
\begin{flushleft}
{\bf Figure 1 $|$ Overview of the jet observed by SDO.} {\bf a-b}, HMI line-of-sight (LOS)
magnetograms displaying the magnetic field environment of the jet footpoint. The green box
in {\bf a} outlines the field-of-view (FOV) of {\bf b} to {\bf f}. The red contour in {\bf b}
represents the footpoint of the jet, which is determined from AIA observations in {\bf d}.
{\bf c-e}, AIA 131 {\AA}, 171 {\AA} and 304 {\AA} images displaying the appearance of the
jet at its peak evolution time. The green box in {\bf e} outlines the FOV of Fig. 3. {\bf f} plots
the temperature derived from the wavelengths of 94 {\AA}, 131 {\AA}, 171 {\AA}, 193 {\AA},
211 {\AA} and 335 {\AA}. The black arrows indicate the hottest temperature patches (HTPs)
in the jet.
\label{fig1}
\end{flushleft}
\end{figure}

\clearpage

\begin{figure}
\centering
\includegraphics
[bb=73 114 486 751,clip,angle=0,scale=0.55]{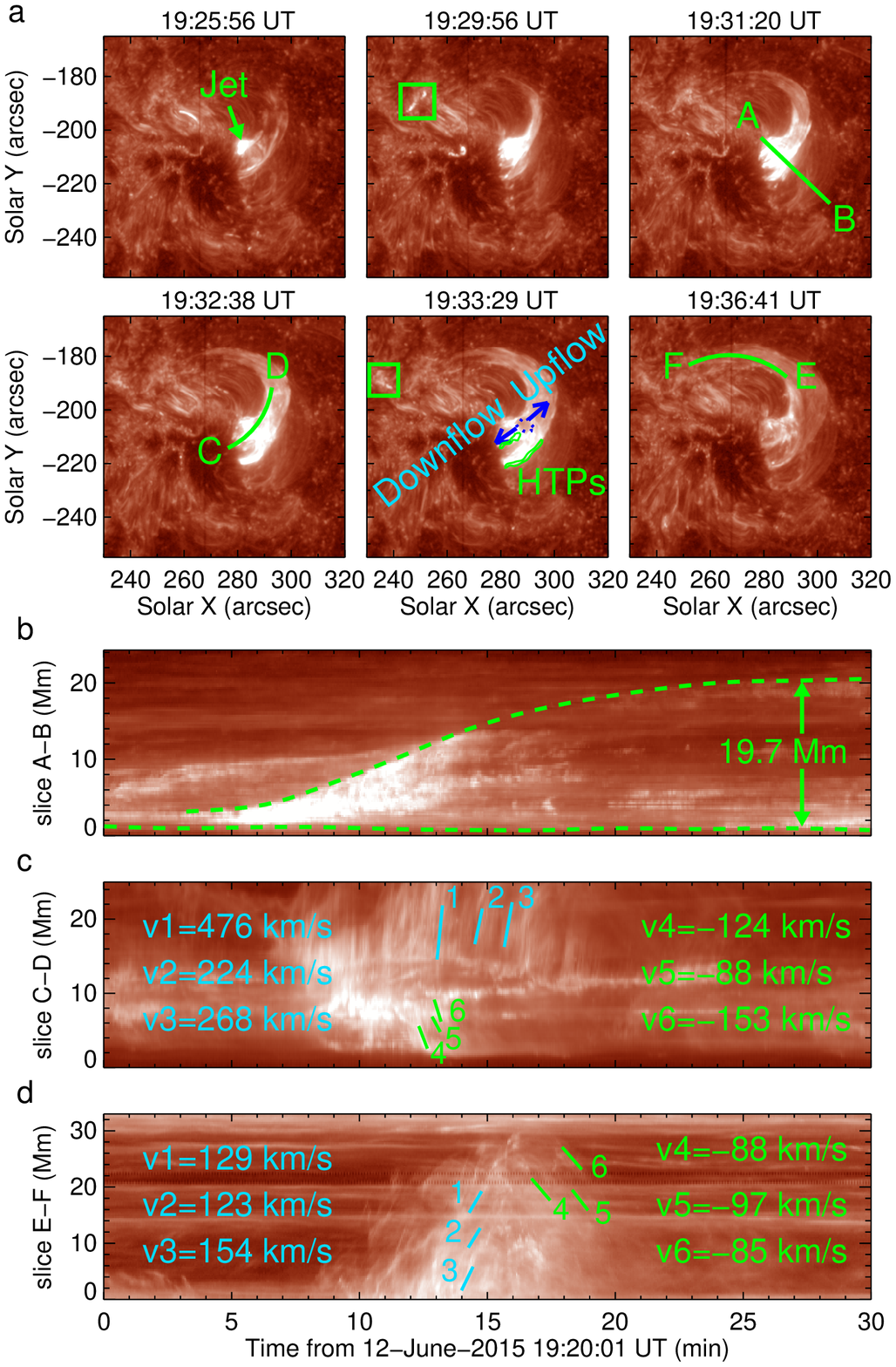}
\begin{flushleft}
{\bf Figure 2 $|$ Evolution of the jet observed by IRIS 1400 {\AA}.}
{\bf a}, IRIS 1400 {\AA} slit-jaw-images (SJIs) displaying the development of the jet.
The green boxes in the second and fifth panels outline the brightenings. The blue
dotted curve in the fifth panel denotes a ``cavity" within the jet. The green line AB
shows the cross-cut position used to obtain the stack plot which displays the jet's
width variation over time, as shown in {\bf b}. The green dashed lines denote the outer
boundary of the jet that reached 19.7 Mm in height by the end. {\bf c}, Evolution at
the position of curve CD (see panel at 19:32:38 UT in {\bf a}) which passes
through the cavity and the bi-directional flows (see panel at 19:33:29 UT in {\bf a}).
The velocities of selected representative bright structures are displayed. {\bf d},
Temporal evolution of flows at the position of curve EF (see panel at 19:36:41 UT in {\bf a}).
The flows from right to left (1$-$3 in cyan color) and from left to right (4$-$6 in green color)
are shown in this plot.
\label{fig2}
\end{flushleft}
\end{figure}

\clearpage

\begin{figure}
\centering
\includegraphics
[bb=60 260 500 560,clip,angle=0,scale=1.0]{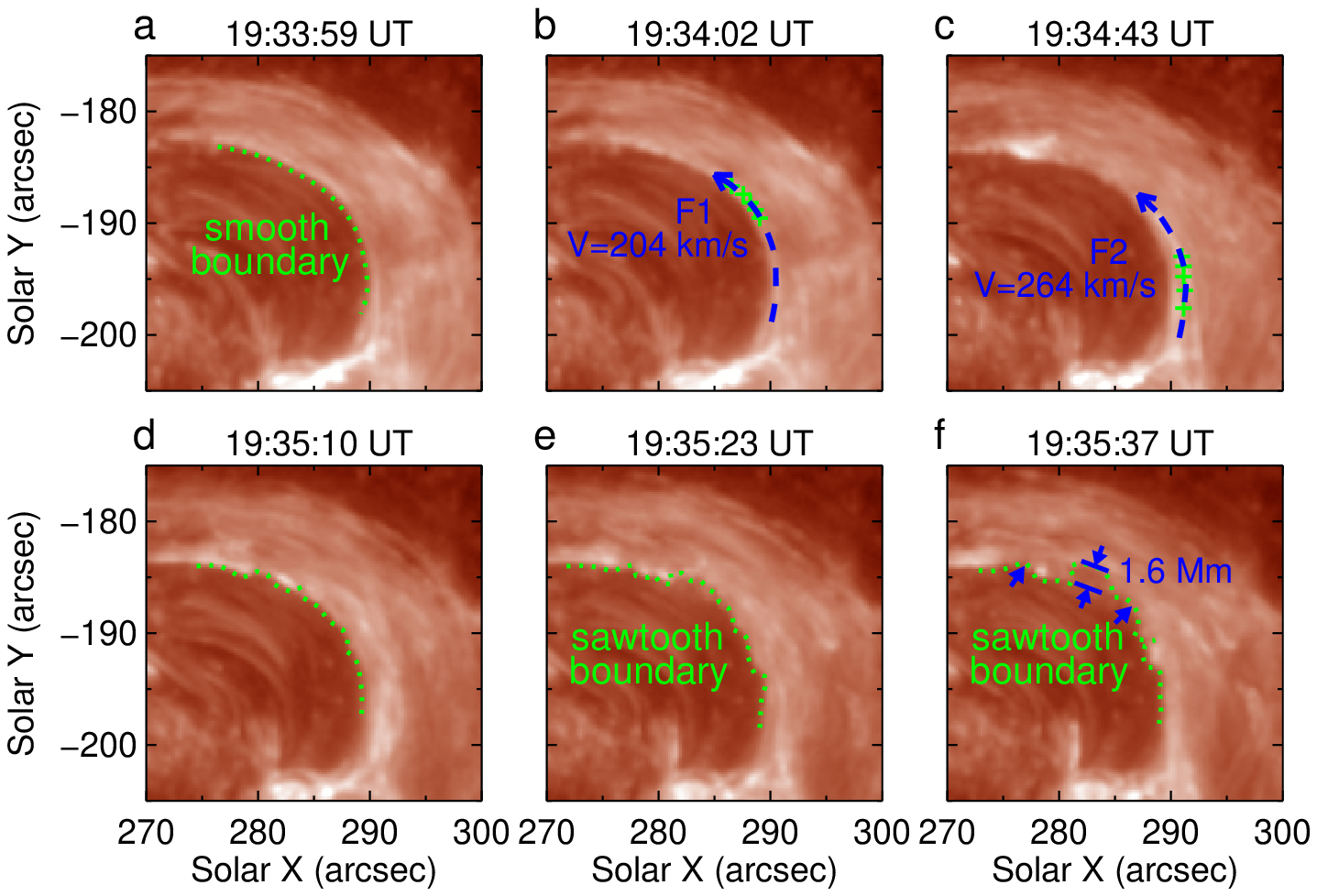}
\begin{flushleft}
{\bf Figure 3 $|$ IRIS 1400 {\AA} SJIs showing the KHI detected at the left boundary of the jet.}
The green curves in panels {\bf a} and {\bf d-f} denote the left boundary which changes from
being smooth ({\bf a}) into a sawtooth pattern ({\bf d-f}). The blue curves
in {\bf b} and {\bf c} denote the trajectories of the first (F1) and the second (F2) flows,
respectively. The blue arrows indicate the directions of ``F1" and ``F2." The green crosses
in {\bf b} ({\bf c}) show the trajectory of a bright point in ``F1" (``F2") which we track to
determine the velocity of ``F1" (``F2"), with the value of 204 km s$^{-1}$ (264 km s$^{-1}$).
The blue arrows in {\bf f} display the distortions of the boundary, with the largest distortion
being 1.6 Mm.
\label{fig3}
\end{flushleft}
\end{figure}

\clearpage

\begin{figure}
\centering
\includegraphics
[bb=10 455 576 740,clip,angle=0,scale=0.8]{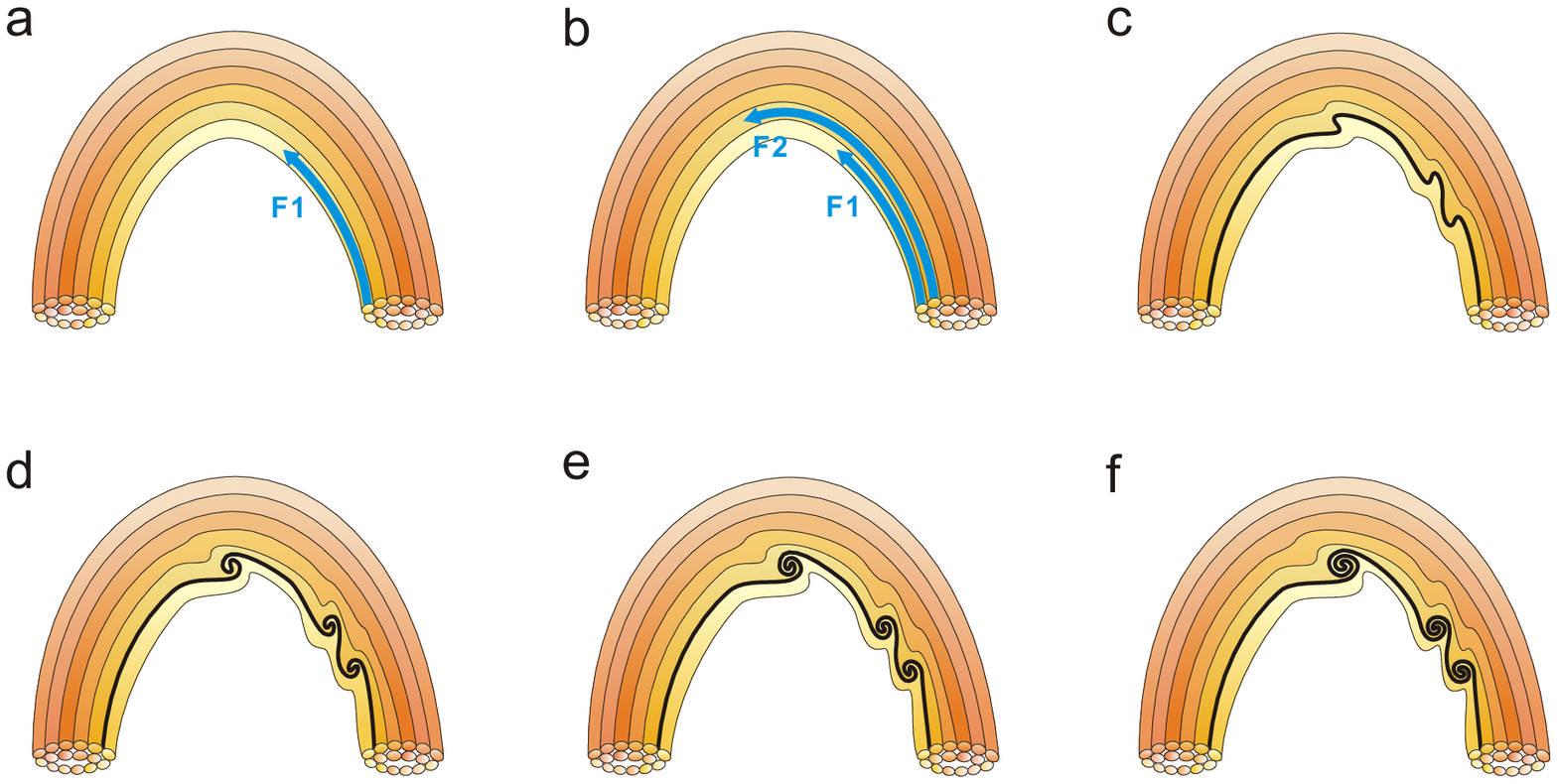}
\begin{flushleft}
{\bf Figure 4 $|$ The development of the KHI in the flux tubes of the jet.} Individual flux tubes are
distinguished with different colors. The skyblue arrows in panels {\bf a} and {\bf b} represent
the flows in the flux tubes. Panels {\bf c-f} show the change of the boundary and the development
of vortices at the interface after the KHI developed.
\label{fig4}
\end{flushleft}
\end{figure}

\clearpage

\begin{flushleft}
{\bf METHODS}

{\bf Instrumentation and data.} 1400 {\AA} SJIs are obtained from the IRIS spacecraft
with a cadence of 3 seconds, and a pixel scale of 0.$\arcsec$33. For the Doppler velocity
measurement, we use data from spectrograph on IRIS in 1402.77 {\AA}. In addition to this, EUV images
from the AIA on board the SDO are employed to display the dynamics of the jet. The
cadence is 12 s and the spatial sampling is 0.$\arcsec$6 pixel$^{-1}$. We use data observed
in 304 {\AA}, 171 {\AA}, 193 {\AA}, 211 {\AA}, 131 {\AA}, 335 {\AA} and 94 {\AA},
which have strong responses to logarithmic temperatures of about 4.7, 5.8, 6.2, 6.3, 7.0, 6.4
and 6.8 Kelvin, respectively. The LOS magnetograms with 45-sec cadence and 0.$\arcsec$5 per
pixel from HMI on SDO are also applied in order to study the magnetic evolution.

{\bf Standard jet model.} The standard jet model was put forward some decades ago$^{38}$.
The key idea of this model is that a current sheet is formed as the jet-base magnetic
arch approaches the ambient opposite-polarity open field. Magnetic reconnection
takes place when the current sheet becomes sufficiently strong and thin.
The current sheet is demolished due to a burst of interchange reconnection between
the jet-base magnetic arch and its ambient field. Magnetic reconnection releases energy
and the plasma in the reconnected field is heated to several million degrees,
hot enough to emit X-rays. Some hot plasma may escape along the reconnected open field
lines and may become the spire of the jet. Meanwhile, a new magnetic arch is formed and may shrink
downward with the reconnection-heated hot plasma in it, which is observed as a bright point in
X-ray images$^{39,40}$.

{\bf Blowout jet model.} According to the blowout jet model$^{26}$, the jet-base magnetic
arch approaches the ambient opposite-polarity open field with intense shear and twist at
the core field in the base arch. The strong shear and twist of the magnetic field in the
core of the jet's base arch drive an ejective eruption, generating blow-out jets and associated
waves$^{28}$. The onset of blowout jets is caused by a burst of reconnection at the interface current sheet,
in the same manner as for standard jets. Then, the sheared-core arch begins to erupt and
drives more breakout reconnections. Reconnection may occur at three different locations: at
the interface current sheet, between the opposite-polarity stretched legs of the internal
erupting field, and, between the leg of the erupting core-field flux rope and its opposite-polarity
ambient open field. The blow-out eruptions continue to build the jet's bright point,
may generate a flare arcade inside the base arch, and may result in a spire with complex
and highly magnetically stressed structure$^{41-44}$. Based on the blowout model, a cartoon is drawn to
display the reconnection scenario of the jet in the text (see Supplementary Fig. 6).
The first drawing is the magnetic topological structure before the jet. The
ambient field is complex, and only a few representative field lines are drawn with yellow
color in order to visualise the connections. The red pattern represents the cold filament material.
New magnetic flux, depicted with blue lines, are emerging constantly and approaching the ambient
field. Reconnections occur at different places and the magnetic structure after the onset of
the jet is shown in Supplementary Fig. 6b. The filament is activated and is ejected out as
displayed by the red contour. After the onset, the reconnections continue to take place and
drive the spectacular evolution of the jet.

{\bf Untwisting in the solar jets.} The helical structure, or spinning, is a common geometrical
feature of solar jets and many studies reveal that it is driven by magnetic untwisting$^{45}$.
In a blowout jet, there is a twisted flux rope (or filament) in the jet's base and the twist may
be released by the reconnection between the flux rope and ambient open fields, thus leading to the
magnetic untwisting$^{46}$. Here, the rotation of the jet may be caused by the magnetic untwisting
as well. The pitch angle is about $30^{\circ}-60^{\circ}$. Assuming that the plasma moves along
the magnetic line and the trajectory represents the morphology of the magnetic field, we estimate
that the ratio of the azimuthal component with regard to the component along the magnetic axis
is approximately 0.6$-$1.7.

{\bf KHI in the solar atmosphere.} KHI was first observed at
quiescent prominences in the solar atmosphere by using data obtained by the Solar
Optical Telescope (SOT) on-board Hinode$^{17,18}$. With the observations by SDO/AIA, KHI
has been detected at the interface between a dimming area and the surrounding corona$^{47}$ and
on the flank of a fast CME less than 150 Mm above the solar surface$^{15}$.
KHI is mostly identified by the appearance of growing ripples or the associated
plume head vortex. The kink-like oscillation of a streamer$^{48}$ and rapid redshifted
and blueshifted excursions$^{49}$ in the solar chromosphere are presumed to be caused by
the KHI as well.

{\bf Magnetic flux estimation.} We estimate the magnetic flux at the footpoint
of the jet using photospheric magnetic field obtained by HMI. According to the
conservation of magnetic flux within the same group of magnetic loops, one can work out the
magnetic flux density of other positions of the jet in case of knowing its area. We assume
that the body of the jet is a cylinder, so its cross-sectional area is easy to obtain since
we can measure the diameter.

{\bf Velocity measurement.} In order to derive the velocities of the bulk flows, we choose some
representative distinct bright points in the flows and trace their movement. The cadence allows
us to measure the distance moved every 3 seconds and the total time and total distance
can be determined easily. Dividing distance by time, one obtains the average velocity.

{\bf Temperature measurement.} Using the differential emission measure (DEM) analysis
method which is based on the ``xrt\_dem\_iterative2.pro'' in the Solar Software
package$^{50,51}$, we derive the temperature map from the AIA 94 {\AA} (Fe XVIII),
131 {\AA} (Fe VIII, Fe XX, Fe XXIII), 171 {\AA} (Fe IX), 193 {\AA} (Fe XII, Fe XXIV),
211 {\AA} (Fe XIV) and 335 {\AA} (Fe XVI) observations. As there is little temperature
discrimination at the low temperatures in the AIA channels, we have to mention that
this method for temperature measurement is only applicable for coronal structures hotter
than $\sim$ 1 MK. In this letter, the jet is visible simultaneously in
multiple AIA channels, which implies that the structure is either broadly multi-thermal,
or at a cool ($\sim$ 10$^5$) temperature$^{52,53}$. Since all AIA channels and the IRIS
channel have a cool-temperature response due to cool, relatively weak spectral lines present
in the passbands, and the jet is also clearly observed in the IRIS 1400 {\AA} passband,
it is more likely that the jet is Transition Region structure. Also, there exists much
cooler filament material in the jet, so the DEM method may bring large error when calculating
the temperature. Up to now, there is no better temperature measurement method that one can use,
and the temperature value provided here is just for reference.

{\bf Spectral analysis.} We employ the IRIS raster level 2 data, which
have been dark corrected, flat fielded, and geometrically corrected, to measure the
associated Doppler shifts. We mainly analyze the emission of the Si IV 1402.77 {\AA} line,
which is formed in the middle transition region with a temperature of $\sim$ 0.065 MK.
The nearby S I 1401.51 {\AA} line is used for the absolute wavelength calibration$^{54}$.
We measure the Doppler shifts of the S I line for seven times (around the moment we focus)
and compare the values with the standard one. The average different value is regarded as
calibration value and the standard deviation is calibration error. To obtain the Doppler shift of the
jet material, we apply a single-Gaussian fitting to the Si IV line profile (see Supplementary Fig.4).
At 19:34:12 UT, a bulk flow from the jet's base to the tail emerged across the slit, as shown
in Supplementary Fig. 4a. By analyzing the spectral profiles of Si IV 1402.77 {\AA}, we find
the Doppler redshift relative to the background was 21.8 km s$^{-1}$ (Supplementary Fig. 4b, c).
The wavelength calibration error is 4.4 km s$^{-1}$. Using the same method, let us now
analyze the motion of the backflow as it passed through the slit (see Supplementary Fig. 4d$-$f).
The backflow progressed along the opposite direction to the original bulk flow and
showed a clear Doppler blueshift, corresponding to an associated
velocity of $-$10.3 km s$^{-1}$. The wavelength calibration error is 4.8 km s$^{-1}$.
The calibration error would change the Doppler velocity value of the flow.
Comparing the calibration error with the Doppler velocity, we conclude that whether
the flow is redshifted or blueshifted won't be changed.

{\bf Theoretical consideration.} As mentioned above, the jet is assumed to be a group of thin
magnetic flux tubes, and ``F1" and ``F2" are the flows in two adjacent flux tubes, respectively.
In the following, we also denote by ``F1" and ``F2" the two flux tubes with bulk flows and
their properties are distinguished with subscript ``1" and ``2". To model the rather complex
situation, we assume the plasma is uniform, incompressible, ideal and
we only consider 2-dimension motion of the flux tube. At the footpoint of ``F1" (``F2"), the
magnetic field intensity is about 434 G (329 G) and is derived from the HMI LOS magnetogram. The flux
tubes have diameters of 100 km at the photosphere. At the location where the KHI developed, the widths
of the flows which have been measured above (630 km for ``F1" and 460 km for ``F2") can be interpreted
as the diameters of the flux tubes. According to the conservation of magnetic flux, therefore,
the magnetic field intensities of ``F1"  and ``F2" are approximately $B_1$ = 11 G and $B_2$ = 15 G.
Using the DEM method, we obtain the total emission measure ($EM$) and the particle number
density $n$ can be derived using $n$ = $\sqrt{EM/l}$ where the $l$ is the depth of the flux tube.
Assuming that $l$ = 500 km, the total $EM_1$ of 8.1 $\times$ 10$^{28}$ cm$^{-5}$ corresponds to a
number density ($n_1$) of 4.0 $\times$ 10$^{10}$ cm$^{-3}$ and the total $EM_2$ of 1.0 $\times$ 10$^{29}$ cm$^{-5}$ corresponds to a number density ($n_2$) of 4.5 $\times$ 10$^{10}$ cm$^{-3}$. Let us suppose that
the plasma density in each tube is homogeneous, the KHI occurs if
\begin{eqnarray}
(\vec{k} \cdot \vec{V_1} - \vec{k} \cdot \vec{V_2})^2 > (\rho_1 + \rho_2)[(\vec{k} \cdot \vec{B_1})^2 + (\vec{k} \cdot \vec{B_2})^2]/\mu_0 \rho_1 \rho_2
\end{eqnarray}
where $\vec{k}$, $\vec{V}$, $\vec{B}$, $\rho$ are the wave vector, velocity, magnetic intensity and mass density in the flux tube, respectively.
Presuming that $\vec{k}$ $\parallel$ $\vec{V_1}$ $\parallel$ $\vec{V_2}$ $\parallel$ $\vec{B_1}$ $\parallel$ $\vec{B_2}$,
then we can work out that the velocity difference threshold is
\begin{eqnarray}
\vartriangle V = \mid \vec{V_1} - \vec{V_2} \mid = \sqrt{(\rho_1 + \rho_2)(B_1 ^2 + B_2 ^2)/\mu_0 \rho_1 \rho_2}.
\end{eqnarray}
Since $\rho$ = $n$ $\cdot$ $m$ ($m$ = 1.673 $\times$ 10$^{-27}$ kg), and $B$ and $n$ have been already
estimated, we obtain that the threshold $\vartriangle$$V$ = 279 km/s. As seen from the formula, the effect
of the magnetic field on the KHI depends on its orientation. The magnetic field component parallel
to the interface discontinuity can exert a restoring force, and can suppress the growth of the KHI, while the
perpendicular component makes no difference$^{55}$. Here, the magnetic field is not strictly parallel
to the flow, as we observe that the second flow rotates as it passes though the side of the jet.
Theoretically speaking, in this case, the KHI will occur as long as the angle between the flow and
magnetic field is more than $43^{\circ}$, which is possible in our situation. These hypotheses are
ideal and the results highly depend on the angle between the flow and magnetic field. However, the actual
situation is complex and the twisting flux tube may lower the threshold of the angle.
Recent three-dimensional (3-D) magnetohydrodynamics (MHD) simulations have indicated that although the draping of
strong tailward component of the Parker-Spiral Interplanetary Magnetic Field tends to stabilize the growth
of the instabilities, KHI with a titled $\vec{k}$ oriented approximately $41^{\circ}$ from the shear
flow plane still develops into the nonlinear phase and the magnetic reconnection may occur as a secondary
effect$^{56}$. The simulations help to understand how the KHI happens indeed for solar jets.

{\bf Growth rate of the KHI.} Once the KHI takes place, we choose two positions (see Fig. 3f where the distortion is
the largest and the place on its right indicated with blue arrow) on the jet and measure these
distortions. The distortions over time are plotted in Supplementary Fig. 6. The distortions are
exponential growths and the growth rates are about 0.059 and 0.067.
In theory, the growth rate is
\begin{eqnarray}
\nonumber
\gamma & = & \sqrt{\rho_1 \rho_2(\vec{k} \cdot \vec{V_1} - \vec{k} \cdot \vec{V_2})^2 - (\rho_1 + \rho_2)[(\vec{k} \cdot \vec{B_1})^2 + (\vec{k} \cdot \vec{B_2})^2]/\mu_0 } / (\rho_1 + \rho_2)     \\
& = & 2\pi \sqrt{\rho_1 \rho_2(\vartriangle V)^2 - (\rho_1 + \rho_2)(B_{1\parallel}^2 + B_{2\parallel}^2)/\mu_0 } / \lambda (\rho_1 + \rho_2).
\end{eqnarray}
Assuming that the angle between the flow and magnetic field is about $45^{\circ}$, with regard the distance
between two vortices as the associated characteristic wavelength ($\sim$ 5000 km), and substituting the value
of the velocity difference into the formula, we can estimate the theoretical growth rate $\gamma \sim$ 0.033. As
the angle between the flow and
magnetic field increases to $60^{\circ}$, the growth rate will rise to 0.093. Remarkably, the measured values
are of the same order of the theoretically estimated values. The deviation may be caused by many reasons. In addition to the measurement error, the assumptions in the theory will also be accountable for the difference.

{\bf Plasma heating of the KHI.}
We study the temperature variation and find that a heating process appears during the KHI as
shown in Supplementary Figure 4a. We choose an area (outlined by the black curve) and calculate
the average temperature, which is about 2 MK hotter during the KHI than before and after the event.
Theoretical studies have shown that the KHI on small scales can play an important role in the
energy dissipation of the associated waves and jets, and in the plasma heating in the solar atmosphere.
Here the temperature increase is thought to be caused by plasma heating process, which is induced
by the KHI. The plasma heating associated with the KHI has been studied before in other
astrophysical phenomena. Combining Cluster observation and MHD numerical simulations, a recent
study provides evidence that ion-scale (200$-$2000 km) fast magnetosonic waves, generating inside the
fluid-scale (36000 km) KH vortices at the Earth¡¯s magnetospheric boundary, have sufficient energy to provide
significant ion heating (2 keV = 20 MK increase for the observed ion flux population), thus demonstrating how
the cross-scale energy transfer from fluid to ion scales can result in ion heating$^{57}$. This cross-scale
mechanisms may also contribute to the heating of the solar corona and play a role in other astrophysical
plasmas$^{37}$. Further, nonlinear 3-D MHD simulations have shown that the resonant dissipation layers are
subject to the KHI, and these resonant layers generate small scale turbulent motions that enhance
the dissipation parameters e.g. eddy viscosity, leading to the dissipation of energy and ultimately
causing heating of the solar corona by Alfv$\acute{e}$n waves$^{58,59}$. This scenario is a rather
viable heating mechanism. The exact heating mechanism in ours reported here needs to be further
examined at a much higher spatial and temporal resolution.

{\bf Data availability.} All the data used in the present study are publicly available.
The IRIS data that support the findings of this study are available from http://iris.lmsal.com/.
The SDO/HMI LOS magnetograms and the SDO/AIA EUV images can be downloaded from http://jsoc.stanford.edu/.

\end{flushleft}

\begin{flushleft}

{\bf References}

1. Lord Kelvin (William Thomson).
Hydrokinetic solutions and observations.
\textit{Philosophical Magazine} \textbf{42}, 362$-$377 (1871).

2. Helmholtz, H. L. F.
\"Uber discontinuierliche Fl\"ussigkeits-Bewegungen [On the discontinuous movements of fluids].
\textit{Monatsberichte der K\"oniglichen Preussische Akademie der Wissenschaften zu Berlin} \textbf{23}, 215$-$228 (1868).

3. Lau, Y. Y. \& Liu, C. S.	
Stability of shear flow in a magnetized plasma.
\textit{Phys. of Fluids} \textbf{23}, 939$-$941 (1980).

4. Ershkovich, A. I.	
Kelvin$-$Helmholtz instability in type-1 comet tails and associated phenomena.
\textit{Space Sci. Rev.} \textbf{25}, 3$-$34 (1980).

5. Begelman, M. C., Blandford, R. D. \& Rees, M. J.
Theory of extragalactic radio sources.
\textit{Rev. Mod. Phys.} \textbf{56}, 255$-$351 (1984).

6. Lobanov, A. P. \& Zensus, J. A.
A cosmic double helix in the archetypical quasar 3C273.
\textit{Science} \textbf{294}, 128$-$131 (2001).

7. Price, D. J. \& Rosswog, S.
Producing ultrastrong magnetic fields in neutron star mergers.
\textit{Science} \textbf{312}, 719$-$722 (2006).

8. Farrugia, C. J. et al.
Charts of joint Kelvin$-$Helmholtz and Rayleigh$-$Taylor instabilites at
the dayside magnetopause for strongly northward interplanetary magnetic field.
\textit{J. Geophys. Res.} \textbf{103}, 6703$-$6727 (1998).

9. Slavin, J. A. et al.
MESSENGER observations of extreme loading and unloading of Mercury's magnetic tail.
\textit{Science} \textbf{329}, 665$-$668 (2010).

10. Hasegawa, A. et al.
Transport of solar wind into Earth's magnetosphere through rolled-up
Kelvin$-$Helmholtz vortices.
\textit{Nature} \textbf{430}, 755$-$758 (2004).

11. Masters, A. et al.
Cassini observations of a Kelvin$-$Helmholtz vortex in Saturn's outer magnetosphere.
\textit{J. Geophys. Res.} \textbf{115}, A07225 (2010).

12. Heyvaerts, J. \& Priest, E. R.
Coronal heating by phase-mixed shear Alfv$\acute{e}$n waves.
\textit{Astron. Astrophy.} \textbf{117}, 220$-$234 (1983).

13. Foullon, C., Verwichte, E., Nakariakov, V. M., Nykyri, K. \& Farrugia, C. J.
Magnetic Kelvin$-$Helmholtz Instability at the Sun.
\textit{Astrophys. J.} \textbf{729}, L8 (2011).

14. Nykyri, K. \& Foullon, C.
First magnetic seismology of the CME reconnection outflow layer in the low corona with 2.5-D MHD
simulations of the Kelvin$-$Helmholtz instability.
\textit{Geophys. Res. Lett.} \textbf{40}, 4154$-$4159 (2013).

15. M\"ostl, U. V., Temmer, M. \& Veronig, A. M. 	
The Kelvin$-$Helmholtz instability at coronal mass ejection boundaries in the solar
corona: observations and 2.5D MHD simulations.
\textit{Astrophys. J.} \textbf{766}, L12 (2013).
	
16. G$\acute{o}$mez, D. O., DeLuca, E. E. \& Mininni, P. D.	
Simulations of the Kelvin-Helmholtz instability driven by coronal mass
ejections in the turbulent corona.
\textit{Astrophys. J.} \textbf{818}, 126 (2016).

17. Berger, T. E. et al.	
Quiescent prominence dynamics observed with the Hinode solar optical telescope. I. turbulent upflow plumes.
\textit{Astrophys. J.} \textbf{716}, 1288$-$1307 (2010).

18. Ryutova, M., Berger, T., Frank, Z., Tarbell, T. \& Title, A.
Observation of plasma instabilities in quiescent prominences.
\textit{Sol. Phys.} \textbf{267}, 75$-$94 (2010).

19. Raouafi, N. E. et al.
Solar coronal jets: observations, theory, and modeling.
\textit{Space Sci. Rev.} \textbf{201}, 1$-$53 (2016).

20. De Pontieu, B., Erd\'elyi, R. \& James, S. P.
Solar chromospheric spicules from the leakage of photospheric oscillations and flows.
\textit{Nature} \textbf{430}, 536$-$539 (2004).

21. Shibata, K. et al.
Chromospheric anemone jets as evidence of ubiquitous reconnection.
\textit{Science} \textbf{318}, 1591 (2007).

22. Tian, H. et al.
Prevalence of small-scale jets from the networks of the solar transition region and chromosphere.
\textit{Science} \textbf{346}, 1255711 (2014).

23. Yokoyama, T. \& Shibata, K.
Magnetic reconnection as the origin of X-ray jets and H$\alpha$ surges on the Sun.
\textit{Nature} \textbf{375}, 42$-$44 (1995).

24. Sterling, A. C., Moore, R. L., Falconer, D. A. \& Adams, M.
Small-scale filament eruptions as the driver of X-ray jets in solar coronal holes.
\textit{Nature} \textbf{523}, 437$-$440 (2015).

25. Chae, J., Qiu, J., Wang, H. \& Goode, P. R.
Extreme-ultraviolet jets and H$\alpha$ surges in solar microflares.
\textit{Astrophys. J.} \textbf{513}, L75$-$L78 (1999).

26. Moore, R. L., Cirtain, J. W., Sterling, A. C. \& Falconer, D. A.
Dichotomy of solar coronal jets: standard jets and blowout jets.
\textit{Astrophys. J.} \textbf{720}, 757$-$770 (2010).

27. Li, X., Yang, S., Chen, H., Li, T. \& Zhang, J.
Trigger of a blowout jet in a solar coronal mass ejection associated with a flare.
\textit{Astrophys. J.} \textbf{814}, L13 (2015).

28. Morton, R. J., Srivastava, A. K. \& Erd\'elyi, R.
Observations of quasi-periodic phenomena associated with a large blowout solar jet.
\textit{Astron. Astrophy.} \textbf{542}, A70 (2012).

29. De Pontieu, B. et al.
The Interface Region Imaging Spectrograph (IRIS).
\textit{Sol. Phys.} \textbf{289}, 2733$-$2779 (2014).

30. Schou, J. et al.
Design and ground calibration of the Helioseismic and Magnetic Imager (HMI) instrument
on the Solar Dynamics Observatory (SDO).
\textit{Sol. Phys.} \textbf{275}, 229$-$259 (2012).

31. Lemen, J. R. et al.
The Atmospheric Imaging Assembly (AIA) on the Solar Dynamics Observatory (SDO).
\textit{Sol. Phys.} \textbf{275}, 17$-$40 (2012).

32. Pesnell, W. D., Thompson, B. J. \& Chamberlin, P. C.
The Solar Dynamics Observatory (SDO).
\textit{Sol. Phys.} \textbf{275}, 3$-$15 (2012).

33. Innes, D. E., Inhester, B., Axford, W. I. \& Wilhelm, K.	
Bi-directional plasma jets produced by magnetic reconnection on the Sun.	
\textit{Nature} \textbf{386}, 811$-$813 (1997).

34. Schlichenmaier, R., Jahn, K. \& Schmidt, H. U.	
Magnetic flux tubes evolving in sunspots. A model for the penumbral fine structure and the Evershed flow.
\textit{Astron. Astrophy.} \textbf{337}, 897$-$910 (1998).

35. Hasegawa, H. et al.
Single-spacecraft detection of rolled-up Kelvin$-$Helmholtz vortices at the flank magnetopause.
\textit{J. Geophys. Res.} \textbf{111}, A09203 (2006).

36. Miura. A. \& Pritchett. P. L.
Nonlocal stability analysis of the MHD Kelvin$-$Helmholtz instability in a compressible plasma.
\textit{J. Geophys. Res.} \textbf{87}, 7431$-$7444 (1982).

37. Retin\`{o}, A.
Space plasmas: A journey through scales.
\textit{Nat. Phys.} \textbf{12}, 1092$-$1093 (2016).

38. Shibata, K. et al.
Observations of X-ray jets with the Yohkoh soft X-ray telescope.
\textit{Publ. Astron. Soc. Jpn.} \textbf{44}, 173L$-$179L (1992).

39. Canfield, R. C. et al.
H alpha surges and X-ray jets in AR 7260.
\textit{Astrophys. J.} \textbf{464}, 1016 (1996).

40. Moreno-Insertis, F., Galsgaard, K. \& Ugarte-Urra, I.
Jets in coronal holes: Hinode observations and three-dimensional computer modeling.
\textit{Astrophys. J.} \textbf{673}, L211$-$L214 (2008).

41. Pucci, S., Poletto, G., Sterling, A. C. \& Romoli, M.
Physical parameters of standard and blowout jets.
\textit{Astrophys. J.} \textbf{776}, 16 (2013).

42. Young, P. R. \& Muglach, K.
Solar Dynamics Observatory and Hinode observations of a blowout jet in a coronal hole.
\textit{Sol. Phys.} \textbf{289}, 3313$-$3329 (2014).

43. Lee, E. J., Archontis, V. \& Hood, A. W.	
Helical blowout jets in the Sun: untwisting and propagation of waves.
\textit{Astrophys. J.} \textbf{798}, L10 (2015).

44. Panesar, N. K., Sterling, A. C., Moore, R. L. \& Chakrapani, P.
Magnetic flux cancelation as the trigger of solar quiet-region coronal jets.
\textit{Astrophys. J.} \textbf{832}, L7 (2016).

45. Patsourakos, S., Pariat, E., Vourlidas, A., Antiochos, S. K. \& Wuelser, J. P.
STEREO SECCHI stereoscopic observations constraining the initiation of polar coronal jets.
\textit{Astrophys. J.} \textbf{680}, L73 (2008).

46. Pariat, E., Antiochos, S. K. \& DeVore, C. R.
A model for solar polar jets.
\textit{Astrophys. J.} \textbf{691}, 61$-$74 (2009).

47. Ofman, L. \& Thompson, B. J.
SDO/AIA observation of Kelvin$-$Helmholtz instability in the solar corona.
\textit{Astrophys. J.} \textbf{734}, L11 (2011).

48. Feng, L., Inhester, B. \& Gan, W. Q.	
Kelvin$-$Helmholtz instability of a coronal streamer.
\textit{Astrophys. J.} \textbf{774}, 141 (2013).

49. Kuridze, D. et al.
Kelvin$-$Helmholtz instability in solar chromospheric jets: theory and observation.
\textit{Astrophys. J.} \textbf{830}, 133 (2016).

50. Weber, M. A., Deluca, E. E., Golub, L. \& Sette, A. L.
Temperature diagnostics with multichannel imaging telescopes in IAU Symposium,
Vol. 223, Multi-Wavelength Investigations of Solar Activity, eds Stepanov, A. V.,
Benevolenskaya, E. E. \& Kosovichev A. G., 321 (2004).

51. Cheng, X., Zhang, J., Saar, S. H. \& Ding, M. D.
Differential emission measure analysis of multiple structural components of coronal
mass ejections in the inner corona.
\textit{Astrophys. J.} \textbf{761}, 62 (2012).

52. Winebarger, A. R. et al.
Detecting nanoflare heating events in subarcsecond inter-moss loops using Hi-C.
\textit{Astrophys. J.} \textbf{771}, 21 (2013).

53. Tian, H. et al.	
Observations of subarcsecond bright dots in the Transition Region above sunspots
with the Interface Region Imaging Spectrograph.
\textit{Astrophys. J.} \textbf{790}, L29 (2014).

54. Tian, H. et al.
Temporal evolution of chromospheric evaporation: case studies of the M1.1 flare
on 2014 September 6 and X1.6 flare on 2014 September 10.
\textit{Astrophys. J.} \textbf{811}, 139 (2015).

55. Chandrasekhar, S.
\textit{Hydrodynamic and Hydromagnetic Stability}
(Oxford: Clarendon, 1961).

56. Adamson, E., Nykyri, K. \& Otto, A.
The Kelvin$-$Helmholtz instability under Parker-Spiral Interplanetary Magnetic Field
conditions at the magnetospheric flanks.
\textit{Adv. Space Res.} \textbf{58}, 218$-$230 (2016).

57. Moore, T. W., Nykyri, K. \& Dimmock, A. P.
Cross-scale energy transport in space plasmas.
\textit{Nat. Phys.} \textbf{12}, 1164$-$1169 (2016).

58. Ofman, L., Davila, J. M. \& Steinolfson, R. S.
Nonlinear studies of coronal heating by the resonant absorption of Alfv$\acute{e}$n waves.
\textit{Geophys. Res. Lett.} \textbf{21}, 2259$-$2262 (1994).

59. Ofman, L. \& Davila, J. M.
Nonlinear resonant absorption of Alfv$\acute{e}$n waves in three dimensions, scaling laws, and coronal heating.
\textit{J. Geophys. Res.} \textbf{100}, 23427$-$23442 (1995).

\end{flushleft}

\clearpage

\begin{flushleft}

{\bf Acknowledgments} \\
This work is supported by the National Natural Science Foundations of
China (11533008, 11790304, 11673035, 11773039, 11673034, 11790300),
Key Programs of the Chinese Academy of Sciences (QYZDJ-SSW-SLH050),
and the Youth Innovation Promotion Association of CAS (2014043).
The data used are courtesy of the SDO and the IRIS science teams.
IRIS is a NASA small explorer mission developed and operated by LMSAL
with mission operations executed at NASA Ames Research Center and major
contributions to downlink communications funded by the Norwegian Space
Center (NSC, Norway) through an ESA PRODEX contract.
R. E. is grateful to STFC (UK) for the awarded Consolidated Grant No. ST/M000826/1,
The Royal Society for the support received in a number of mobility grants. He also
thanks the Chinese Academy of Sciences Presidents International Fellowship
Initiative, Grant No. 2016VMA045 for support received.

{\bf Author contributions} \\
X. L. deduced and analysed the data, and wrote the manuscript with help from
all authors. J. Z. discovered the original observations. S. Y. and Y. H. contributed
to the data processing. R. E. led the discussions and interpretation. All authors
discussed the results and commented on the manuscript.

{\bf Additional Information} \\
{\bf Supplementary information} is available for this paper.\\
{\bf Reprints and permissions information} is available at www.nature.com/reprints. \\
{\bf Correspondence and requests for materials} should be addressed to
X. L. (lixiaohong@nao.cas.cn) or J. Z. (zjun@nao.cas.cn).

{\bf Competing Interests}\\
The authors declare no competing interests.

\end{flushleft}

\end{document}